\documentclass[manuscript]{aastex61}
\usepackage{txfonts}
\usepackage{latexsym,bm}
\usepackage{multirow}
\shorttitle{MODULATION OF PAMELA GCRS}

\shortauthors{Qin and Shen}

\begin{document}

\title{Modulation of Galactic Cosmic Rays in the Inner Heliosphere, comparing 
with PAMELA measurements}
\correspondingauthor{G. Qin}
\email{qingang@hit.edu.cn}

\author[0000-0002-3437-3716]{G. Qin}
\affiliation{School of Science, Harbin Institute of Technology, Shenzhen, 
518055, China}
\affiliation{State Key Laboratory of Space Weather, National
Space Science Center, Chinese Academy of Sciences,
Beijing 100190, China}
\affiliation{College of Earth Sciences, University of
Chinese Academy of Sciences, Beijing 100049, China}

\author[0000-0002-4148-8044]{Z.-N. Shen}
\affiliation{State Key Laboratory of Space Weather, National
Space Science Center, Chinese Academy of Sciences,
Beijing 100190, China}
\affiliation{College of Earth Sciences, University of
Chinese Academy of Sciences, Beijing 100049, China}

\begin{abstract}

We develop a numerical model to study the time-dependent modulation 
of galactic cosmic rays (GCRs) in the inner heliosphere. In the model 
a time-delayed modified Parker heliospheric magnetic field (HMF) and a new 
diffusion coefficient model, NLGCE-F, from Qin \& Zhang (2014), are adopted. 
In addition, the latitudinal dependence of magnetic turbulence magnitude 
is assumed as $\sim (1+\sin^2\theta)/2$ from the observations of Ulysses, 
and the radial dependence is assumed as $\sim r^S$, where we choose an 
expression of $S$ as a function of the heliospheric current sheet (HCS) 
tilt angle. We show that the analytical expression used to describe the 
spatial variation of HMF turbulence magnitude agrees well with the Ulysses, 
Voyager 1, and Voyager 2 observations.   
By numerically calculating the modulation code we get the proton energy 
spectra as a function of time during the recent solar minimum, it is shown 
that the modulation results are consistent with the PAMELA measurements.

\end{abstract}

\keywords{cosmic rays, modulation, turbulence, activity – Sun: heliosphere}

\section{Introduction}
Galactic cosmic rays (GCRs) are modulated by solar wind irregularities 
while transporting inside the heliosphere. The physical mechanism of cosmic ray 
transport in the heliosphere is described by a well known Parker transport 
equation (TPE) \citep{Parker1965},
\begin{equation}
\frac{\partial f}{\partial t} =-\left(\mathbf{V}_{sw}+ \left\langle{\mathbf{v}_d}\right\rangle
\right) \cdot\nabla f  +\nabla\cdot\left(\mathbf{K}_s\cdot\nabla f \right)+\frac{1}
{3}\left(\nabla\cdot\mathbf{V}_{sw}\right)\frac{\partial f}{\partial\ln p}, 
\label{eq:TPE}
\end{equation} 
where $f(\mathbf{r},p,t)$ is the omni-directional cosmic ray distribution 
function, with $\mathbf{r}$ the position, $p$ the particle momentum 
and $t$ the time. The distribution function $f(\mathbf{r},p,t)$ is 
related to the differential intensity $j$ with respect to kinetic energy 
by $j=p^2f$. The terms on the right hand side include all 
the relevant transport processes, i.e., the outward convection by the solar 
wind velocity $\mathbf{V}_{sw}$, the pitch-angle averaged drift velocity 
$\left\langle{\mathbf{v}_d}\right\rangle$ caused by irregularity in the global 
heliospheric magnetic field, $\mathbf{K}_s$ is the symmetric part of the 
diffusion tensor which is diagnal in HMF-aligned coordinate system, and the adiabatic energy loss, 
which have been successfully illustrated by theoretical and numerical models 
\citep[see, e.g., ][]{Parker1965,Zhang1999,Pei2010,Strauss2012,Potgieter2013,
Zhao2014}.

The past solar cycle (Cycle 24) was unusual with a prolonged periodicity, and the solar 
minimum conditions lasted until early 2010. 
During the solar minimum of 2006 to 2009, the averaged sunspot number (SSN) was reported to 
be the lowest since 1914 \citep{Schrijver2011}. The solar polar field strength during the 
2003-2004 solar maximum was especially weaker than previous three solar cycles
\citep{Svalgaard2005}. The heliospheric magnetic field (HMF) 
reached $\sim$3 nT at 1 AU in 2009, the lowest value since 1963 \citep{Ahluwalia2010}. 
The tilt angle ($\alpha$) of the heliospheric current sheet (HCS) had a flatter decline than 
previous solar minimums \citep{Ahluwalia2011}. It is also reported that the coronal mass 
ejection (CME) 
rates and the solar wind dynamic pressure in 2008-2009 were noticeably lower 
than that in 1997-1998 \citep{Vourlidas2010,McComas2008}. These extreme solar minimum 
conditions resulted 
in a record high-level GCR intensity measured at Earth \citep{Mewaldt2010,Ahluwalia2011}. 
Such a prolonged solar minimum provides a good chance to study the modulation of GCRs inside 
the heliosphere \citep[e.g.,][]{Mewaldt2013,Adriani2013,Potgieter2014,Zhao2014}.

The Payload for Antimatter-Matter Exploration and Light-nuclei Astrophysics (PAMELA)
satellite experiment was designed to study the charged components of cosmic rays (CRs), among 
which antiparticles are focused. PAMELA has been taking data since it was launched in June 
2006, and has brought plentiful scientific results about the heliosphere
\citep{Adriani2014,Mori2015,Galper2017}. PAMELA 
measures precise cosmic ray proton and helium spectra in the rigidity range from $1$ GV to 
$1.2$ TV 
\citep{Adriani2011}. The energy spectra of protons and helium particles show different 
spectral indexes 
above 30 GV and both present a spectral hardening at about $230$ GV. The electron and positron 
energy spectra are measured up to $600$ GeV and $200$ GeV, respectively 
\citep{Adriani2011e, Adriani2013e}. 
PAMELA also measures the flux of Boron and Carbon as well as the boron-to-carbon (B/C) ratio which 
can be utilized to investigate the cosmic ray propagation processes \citep{Adriani2014BC}. 
\citet{Adriani2013} presented precise galactic proton energy spectra in the range $0.08-50$
GeV for 
each Carrington rotation from July 2006 to January 2010. The observed proton spectra became 
progressively 
softer since July 2006. Later, \citet{Adriani2015} reported precise electron spectra with a 
six-month 
interval at the same time period. Such precise CR energy spectra can be used to study physical 
processes of CR transport in the heliosphere, including convection, diffusion, 
drifts, and adiabatic energy changes.

The precise cosmic ray spectra measured by PAMELA instrument provides
important information for people to understand the origin and propagation of GCRs. Voyager 1 was 
assumed to have crossed the heliopause (HP) at $\sim 122$ AU on August 2012 \citep{Webber2013},
cosmic 
ray spectra in the energy between $5$ MeV to $50$ MeV  in the very local interstellar medium was
then reported 
by \citet{Stone2013}. The cosmic ray spectra data from PAMELA and Voyager 1 can be 
used to construct the very local interstellar spectrum
\citep[LIS, see e.g.,][] {Potgieter2014,Ngobeni2014,Potgieter2015,Vos2015,Bisschoff2016}. 
The LIS is important as the input spectrum in the solar modulation study. To 
understand the modulation processes during this unusual solar minimum, three-dimensional 
(3D) numerical 
models have been established to solve the Parker transport equation. \citet{Zhao2014} used 
an empirical 
diffusion coefficient model according to \citet{Zhang1999} and incorporated a 3D wavy HCS. 
By reproducing the proton spectra observed by PAMELA and IMP 8 in previous 
two solar minima, they concluded that increased parallel diffusion and decreased 
perpendicular
diffusion in polar direction caused by low magnetic turbulence might be the possible 
mechanism for the high GCR intensity in the past solar minimum, which is in contrast 
to the assumption of enhanced diffusion in the polar regions that was 
used to explain the observed Ulysses CR gradients \citep[e.g.,][]{Potgieter2000}.
\citet{Potgieter2014} used a different diffusion coefficient model, in which modulation 
parameters vary as a function of time, in their numerical work. They successfully 
reproduced the PAMELA 
proton spectra for four selected periods and reported that this solar minimum was 
``diffusion dominated", and that the modulation effects of particle drifts 
were less obvious but still played a significant role. Similar numerical model was also used to
study the modulation of galactic 
electrons \citep{Potgieter2015} and the modification effects of Parker HMF over the polar region 
\citet{Raath2016}.

Furthermore, due to the precise proton and 
electron spectra measured by PAMELA from 2006 to 2009 \citep{Adriani2013,Adriani2015}, 
one is able 
to know the different modulation effects between protons and electrons, which is called the 
charge-sign-dependent modulation. In this past
polarity $A<0$ solar minimum, low-energy protons were more sensitive to the changes of 
heliospheric 
conditions than low-energy electrons, such phenomenon can be reproduced only by incorporating 
drifts in the numerical model \citep{Felice2017}. The PAMELA proton flux data were also used 
to study the spatial gradients in the inner heliosphere together with proton flux data 
from Ulysses COSPIN/KET. \citet{Simone2011} and \citet{Gieseler2016} used an empirical approach to 
calculate the radial and latitudinal gradients of protons during the past solar minimum. It is shown
that the radial gradients are always positive while the latitudinal gradients are always negative
as expected but with less magnitude than that predicted by earlier works 
\citep{Potgieter2001}. Following \citet{Simone2011} and \citet{Gieseler2016}, \citet{Vos2016} 
also used a numerical model to compute the spatial gradients of protons from 2006 
to 2009. They concluded that although the drift effects were weaker than the predictions from 
those drift-dominated works due
to the suppression by the excess diffusion, they still played an important 
role due to the significant decrease of HMF magnitude until the end of 2009. 

Note that numerical models of modulation are usually solved in steady state, 
interplanetary conditions have to be determined with the solar activity some time before because of
the limit of solar wind speed. 
\citet{Ndiitwani2013} used a time-dependent two-dimensional (2D) numerical model to study 
CR modulation using PAMELA proton data in this unusual period. Smoothed monthly HMF and HCS, which were 
embedded in 
the solar wind plasma, were used to establish a realistic heliospheric conditions. 
Based on the work of \citet{Manuel2011} and \citet{Potgieter2014}, \citet{Ndiitwani2013} 
established a 
time-dependent diffusion coefficient model, in which the yearly time-dependent modulation parameters 
 were obtained from the compound model \citep{Manuel2011} 
and the empirical model \citep{Potgieter2014}, respectively. However, \citet{Ndiitwani2013} did not
consider the variations of solar wind speed. Recently, 
\citet{Boschini2017} used a 2D heliospheric modulation 
\citep[HelMod, e.g., ][]{Bobik2012,Bobik2013} model to study the modulation of GCR 
during solar cycles 23 and 24. In their model, the heliosphere was divided into 
polar and equatorial regions, the modified Parker spiral HMF \citep{Jokipii1989} 
and the Parker spatial HMF \citep{Parker1958} were used in polar and equatorial 
regions, respectively. They also re-scaled the heliosphere into 15 radially 
equally-spaced slices to relate interplanetary conditions with states near Earth. 
They used a parameter $K_0$ to describe the time dependence of diffusion coefficients. 
\citet{Bobik2012} discussed the relationship between $K_0$ and the modulation 
strength given by the force-field model \citep[FFM, see e.g., ][]{Gleeson1968,Gleeson1971}, 
and \citet{Boschini2017} derived $K_0$ using modulation strength data from 
\citet{Usoskin2011}. For periods of low solar activity, it was divided into 
ascending and descending phases for both negative and positive solar magnetic field
polarities, and different polynomial equations were used to describe the relationship 
between $K_0$ and the sunspot numbers. Furthermore, they used neutron monitor 
counting rate to reproduce the variation of $K_0$ during periods of high solar 
activity. Such model was used to study modulation of GCRs with energy 
approximately larger than 0.5 GeV/nucleon, and the results were consistent with the 
observations of PAMELA, AMS-02, and Ulysses.

In this paper, we develop a model of GCR modulation in the inner heliosphere  to study
the GCRs measurements from PAMELA. The paper is organized as follows: In Section 2 
we discuss the GCRs modulation model, including the interplanetary conditions input 
from observations, heliospheric magnetic field and solar
wind speed,  the particle drifts, the magnetic turbulence throughout the 
inner heliosphere, the diffusion  coefficients, and the heliospheric boundary, 
from subsections 2.1 to
2.6. In section 3, we describe the numerical methods. 
In section 4, we show the numerical modulation results and the comparison with the 
PAMELA observations in recent solar minimum. 
Conclusions and discussion are shown in section 5.

\section{GCRs modulation model}
\subsection{Interplanetary conditions input from observations}

\begin{figure}
\epsscale{1.} \plotone{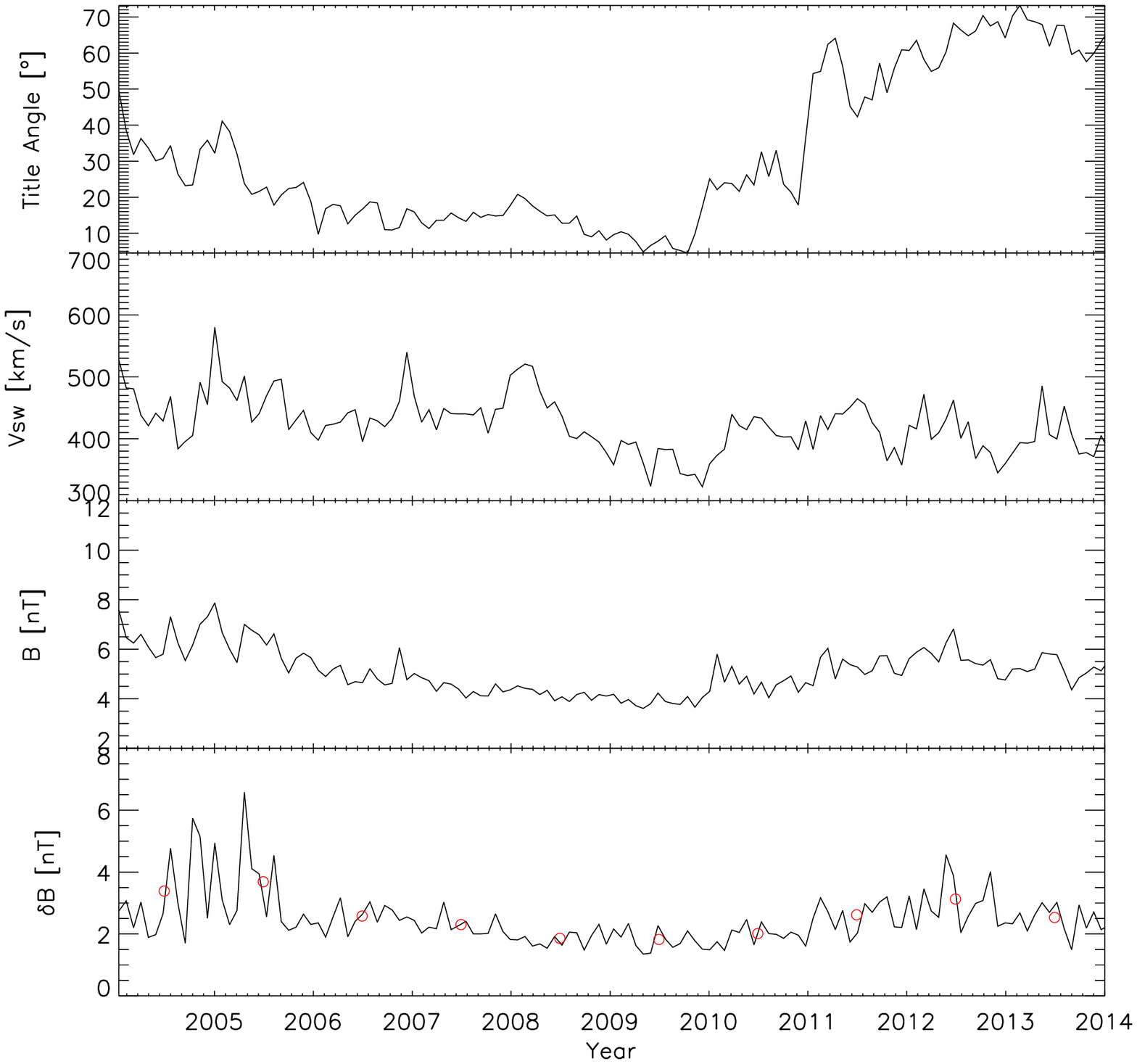}
   \caption{Input interplanetary parameters at 1 AU. Top panel shows the title 
angle of heliospheric current sheet from the WSO website (\url{wso.stanford.edu}) 
with “new” model. Second and third panels represent averaged solar wind velocity 
and averaged magnetic field strength for each Carrington rotation, respectively. 
Black line in the bottom panel means the square root of statistical variance 
$\delta B^2$ which is calculated over Carrington rotation intervals using hourly 
averages of HMF magnitude from OMNI website (\url{omniweb.gsfc.nasa.gov}). Red 
circles represent yearly magnetic turbulence magnitude from \citet{Manuel2014}.}
   \label{fig:input}
 \end{figure}

In order to study the time-dependent modulation of GCRs, we need some 
spacecraft observations near the Earth.  
Figure \ref{fig:input} illustrates observations of interplanetary conditions as 
a function of time which is used in our model. 
Top panel shows the computed tilt angle $\alpha$ until 2015 for the 
new model from Wilcox Solar Observatory (\url{wso.stanford.edu}). 
Second and third panels show averaged solar wind velocity $V_{sw}$ and 
HMF magnitude at $1$ AU using the OMNI data (\url{omniweb.gsfc.nasa.gov}) 
for each Carrington rotation. 
Based on the assumption of isotropic magnetic turbulence, the total 
variance $\delta B^2$ is calculated over Carrington rotation intervals using 
hourly averages of HMF magnitude from OMNI. We have $n$ ($\sim655$ here) samples 
per Carrington rotation, the variance of the total magnetic field magnitude is 
\begin{equation}
\delta B^2=\frac{1}{n}\sum_{i=1}^{n}\left(B_i-\overline{B}\right)^2, 
\end{equation}
where 
\begin{equation}
\overline{B}=\frac{1}{n}\sum_{i=1}^{n}B_i.
\end{equation}
The square root of $\delta B^2$ is shown as black line in the bottom panel of 
Figure \ref{fig:input}. \citet{Manuel2011,Manuel2014} 
\citep[see also, e.g., ][]{Strauss2010,Strauss2011a} also calculated the total 
magnetic field variance over 1 year intervals, and the results are shown as red 
circles in the bottom panel. It is shown that our calculation is consistent with 
the results of \citet{Manuel2011,Manuel2014}.
In the GCRs modulation model, all input parameters are obtained from 
observations near the Earth shown in Figure \ref{fig:input}.

\subsection{Heliospheric magnetic field and solar wind speed}

The heliospheric magnetic field (HMF), which plays an important role in the modulation
of GCRs, is assumed to have an Archimedean 
spiral due to the solar rotation according to \citet{Parker1958}. 
The Parker spiral HMF can be written as 
\begin{equation}
\mathbf{B}=\frac{AB_0}{r^2}\left(\mathbf{e}_r - \frac{(r-r_s)\Omega\sin\theta}
{V_{sw}}\mathbf{e}_\phi\right)\left[1-2H(\theta-\theta')\right],
\label{eq:parker}
\end{equation}
where $B_0$ is a constant, $A$ is the polarity of 
HMF whose positive (negative) value represents the magnetic field points outward 
(inward) in the northern hemisphere of Sun, $\mathbf{e}_r$ and $\mathbf{e}_\phi$ 
are unit vectors in the radial and azimuthal directions, respectively, $r$ is 
heliocentric 
radius, $\theta$ is the polar angle, $r_s$ is the radius 
of the source surface where the HMF is assumed to be directed radially outwards 
and we take $r_s=r_\odot=0.005 \textrm{AU}$ with $r_\odot$ being the radius 
of solar surface \citep{Jokipii1989},  
$\Omega=2.66\times10^{-6}$ rad s${}^{-1}$ is the rotation speed of Sun, 
$V_{sw}$ is the radial solar wind speed, $\theta^\prime$ is the heliospheric 
current sheet (HCS) latitudinal extent, and $H$ is the Heaviside function. 

However, the Parker HMF is an oversimplification and gives a low magnetic field 
intensity at large radial distance in polar heliosphere, which could lead 
to the too rapid entry of GCRs in the polar regions, so it is necessary to modify
the Parker HMF.
\citet{Jokipii1989} suggested superimposing a 
perturbation field on the Parker spiral HMF since turbulence near the solar 
surface resulting in a transverse magnetic field at large radial distance in 
the polar regions. With this modification the Parker spiral HMF becomes 
\begin{equation}
\mathbf{B}=\frac{AB_0}{r^2}\left(\mathbf{e}_r + 
\frac{r\delta(\theta)}{r_s}\mathbf{e}_{\theta} -\frac{(r-r_s)\Omega\sin\theta}
{V_{sw}}\mathbf{e}_\phi\right)[1-2H(\theta-\theta')],
\label{eq:parker1}
\end{equation}
where $\delta(\theta)$ is the perturbation parameter. In order to have a 
divergence free magnetic field the perturbation parameter is written as 
\begin{equation}
\delta(\theta)=\frac{\delta {}_m}{\sin \theta},
\end{equation}
here $\delta_m$ indicates the perturbation parameter in the equatorial plane. 
In addition, we use a reflective boundary condition near the poles to 
avoid singularity, 
$\theta=2\theta_0-\theta$, for $\theta<\theta_0$ if $\theta<90^\circ$ or 
$\theta>\theta_0$ if $\theta>90^\circ$. 
In this study we set $\delta {}_m=2\times 10^{-5}$ \citep{Bobik2013,Boschini2017} 
, $\theta_0= 2.5^\circ$ if $\theta<90^\circ$ and $\theta_0= 177.5^\circ$ if 
$\theta>90^\circ$. 
This modification makes the field decrease as $r^{-1}$ instead of $r^{-2}$ in 
the polar regions for large $r$ without changing the magnetic field dramatically 
in the equatorial plane, and it is supported by the observations of Ulysses 
\citep[e.g., ][]{Balogh1995,Heber2006} and tested by 
numerical models \citep[e.g.,][]{Langner2004, Bobik2012,Bobik2013,Raath2016}.

The solar wind speed has a latitudinal
dependence during solar minimum, increasing from $\sim 400$ km s${}^{-1}$ in 
the equatorial plane to $\sim 800$ km s${}^{-1}$ in the high latitudes, 
but during solar maximum, such simple pattern does not exist anymore 
\citep{McComas2002,Heber2006,Zurbuchen2007}. 
Solar activity can be classified in terms of the HCS tilt angle 
$\alpha$, with $\alpha\leq 30^\circ$, $30^\circ<\alpha\leq 60^\circ$, and 
$60^\circ<\alpha\leq 90^\circ$ representing periods of low, moderate, and high solar 
activity, respectively \citep{Potgieter2001,Potgieter2013a}.
In addition, the solar wind speed accelerates from zero to a constant 
within 0.3 AU from the Sun according to \citet{Sheeley1997}. 
In this work, we study GCRs in solar minimum, so following \citet{Heber2006} and 
\citet{Potgieter2013} we express solar wind speed as 
\begin{equation}
\mathbf{V}_{sw}(r,\theta) = V_0\left\{1-exp\left[\frac{40}{3}\left(\frac{r_s-r}{r_0}\right)\right]
\right\}\left\{1.475\mp 0.4\tanh\left[6.8(\theta-\frac{\pi}{2}\pm\xi)\right]\right\}\mathbf{e}_r, \label{eq:Vsw}\\
\end{equation}
with $V_0=400$ km/s, $r_0=1$ AU, and $\xi=\alpha+15\pi/180$. 
The top and bottom sign correspond to the northern and southern hemisphere, 
respectively. However, if we study GCRs during periods of moderate and high solar
activities the solar wind speed can be set as a constant extracted from OMNI data 
set \citep{Bobik2012}.
Note that for simplicity purpose, in solving the TPE Equation (\ref{eq:TPE}) 
numerically in each step we
assume the magnitude of solar wind as a constant with the value calculated
with Equation (\ref{eq:Vsw}). 

\begin{figure}
\epsscale{0.8} \plotone{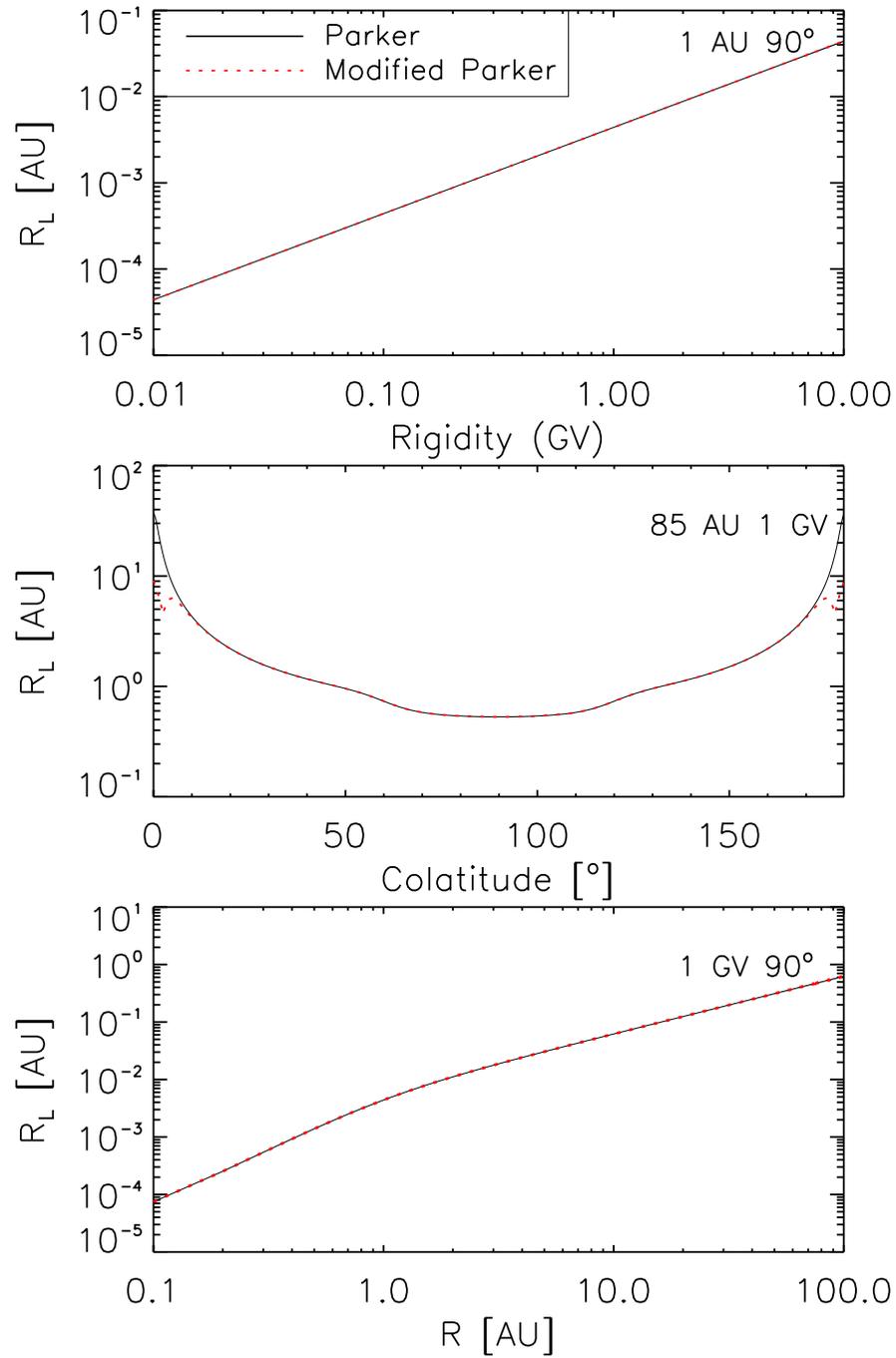}
   \caption{Particle's gyro-radius is shown as functions of rigidity, 
colatitude at $85$ AU, and radial distance in the equatorial plane in top, middle,
and bottom panels, respectively. }
   \label{fig:RL}
 \end{figure}

Figure \ref{fig:RL} shows particle's gyro-radius as a function of rigidity (top 
panel), polar angle (second panel), and radial distance (bottom panel). 
Interplanetary conditions at $1$ AU are set 
as $B=5.05$ nT, and $\alpha=15^\circ$. The black solid and red dotted lines 
indicate results from the Parker field and the modified one, respectively.
From the figure we can see that generally the 
modified field agrees with the Parker field, however, for $1$ GV particles,
in the polar regions with large solar radial distance, very weak Parker field makes 
particles' gyro-radius very large, but the modified model with enhanced field
keeps particles' gyro-radius around several AU.

It is noted that interplanetary conditions at solar radial distance
$r$ are related to the states at the source surface $r_s$ at some earlier 
time because of the solar wind flow \citep{Potgieter2014,Potgieter2015}, and the 
heliosphere is dynamic due to the solar activities. In our numerical model, 
we divide time in days and assume a locally static heliosphere in each day. 
In the $i^{th}$ Carrington period $t_i$, the observation of solar wind 
velocity at $1$ AU is $v_i$. 
To calculate the interplanetary conditions in solar distance $r$ at time $t$, we can use
the input parameters near the Earth (e.g., $V_{sw}$, $B$, $\delta B$, $\alpha$, $A$) at 
time $t_i$ if Equation (\ref{eq:SVSW}) is satisfied,
\begin{equation}
\left\{
\begin{array}{l}
v_i(t-t_i) \geq r-r_0 \\
v_{i+1}(t-t_{i+1}) < r-r_0 \\
\end{array}
\right .
\label{eq:SVSW}
\end{equation}
with $r_0=1$ AU.
For simplification, we use 
\begin{equation}
\left\{
\begin{array}{l}
v_0(t-t_i) \geq r-r_0 \\
v_0(t-t_{i+1}) < r-r_0. \\
\end{array}
\right .
\end{equation}
with the typical solar wind speed $v_0=0.25$ AU/day. It is noted that in the region
between two slices of plasma the HMF is not divergence free, but in each step of the
numerical solution of the TPE Equation (\ref{eq:TPE}), we always keep inside one slice of
plasma.

\subsection{Particle drifts}

Particle drifts play an important role in the solar modulation of GCRs 
\citep{Jokipii1977,Jokipii1979,Potgieter2013}, 
the pitch angle averaged drift velocity caused by irregularity in the HMF is given 
by
\begin{equation}
\langle \mathbf{v}_d \rangle=\nabla\times\left(\kappa_A\frac{\mathbf{B}}{B}\right),
\end{equation}
with $\kappa_A$ the drift coefficient. Under the assumption of weak scattering, 
the drift coefficient is simply written as 
\begin{equation}
\kappa_A=q\frac{P\beta}{3B}, 
\end{equation}
with q the particle charge sign, $P$ the rigidity of particle and 
$\beta$ the ratio between the speed of 
particle and that of light. For the modified Parker HMF given in Equation 
(\ref{eq:parker1}), the drift velocity can be written as \citep{Burger1989}
\begin{eqnarray}
\langle \mathbf{v}_d \rangle & =&q\frac{P\beta}{3}\nabla\times\left(\frac{\mathbf{B}}
{B^2}\right)\nonumber\\
	& =&qA\frac{P\beta}{3}[1-2H(\theta-\theta^{\prime})]\nabla\times\mathbf{f}\nonumber\\
    &{}&+qA\frac{2\beta P}{3}\delta_{Dirac}(\theta-\theta^\prime)\mathbf{f}\times\nabla(\theta-
	\theta^{\prime})
 \nonumber\\
 & \equiv&\mathbf{v}_{gc}+\mathbf{v}_{ns},\label{eq:drift}
\end{eqnarray}
where 
\begin{eqnarray}
\mathbf{f} &=& \frac{r^2\left(\mathbf{e}_r+
	\eta\mathbf{e}_{\theta} -\Gamma\mathbf{e}_{\phi}\right)}{B_0\left(1+\eta^2+\Gamma^2\right)},\\ 
\eta&=&\frac{r\delta_m}{r_s\sin\theta},\\ 
\Gamma&=&\frac{r\Omega\left(r-r_s\right)\sin\theta}{V_{sw}}.
\end{eqnarray}
Here, $\delta_{Dirac}$ is the Dirac's delta function, $\mathbf{v}_{gc}$ is the 
combination of gradient and curvature drifts, and $\mathbf{v}_{ns}$ is the 
current sheet drift. 

In the following, we show that charge-sign dependent modulation and a 22-year cycle
could be caused by gradient and curvature drifts \citep{Potgieter2013}. 
During $A < 0$ polarity cycles, protons mainly drift inwards  
along the HCS in the equatorial regions so their intensity can be reduced by the 
increasing waviness of HCS, therefore, a sharp peak in the temporal 
profile of GCR intensity is usually observed. However, during the $A > 0$ cycles, 
protons mainly drift inwards 
from polar regions, therefore, a flatter peak of GCR intensities are usually
observed. This effect reverses for negatively charged GCRs. 
The radial, latitudinal, and azimuthal components of the gradient and 
curvature drifts are given by 
\begin{eqnarray}
v_{gc,r}&=& -v_{gc,0}\left(1+2\eta^2\right)\Gamma\cot\theta\label{eq:gcdriftr}\\
v_{gc,\theta}&=& v_{gc,0}\left(2+\eta^2+\Gamma^2\right)\Gamma\label{eq:gcdrifttheta}\\
v_{gc,\phi}&=& v_{gc,0}\left[\eta\left(2+\eta^2+\Gamma^2\right)+
	\left(-\eta^2+\Gamma^2\right)\cot\theta\right],\label{eq:gcdriftphi}
\end{eqnarray}
respectively, where 
\begin{equation}
v_{gc,0}=qA\frac{2P\beta r[1-2H(\theta-\theta^{\prime})]}{3B_0(1+\eta^2+\Gamma^2)^2}. 
\end{equation}
The expression for $\theta^\prime$ is given by \citet{Kota1983}
\begin{equation}
\theta^\prime = \frac{\pi}{2}-\arctan\left[\tan\alpha\sin\left(\phi+
\frac{\left(r-r_s\right)\Omega}{V_{sw}}\right)\right], 
\label{hcs}
\end{equation}
with $\alpha$ the tilt angle. This formula is valid for large tilt angle 
conditions \citep{Pei2012,Raath2015}. 

In the current sheet, the current sheet drift velocity given by 
Equation (\ref{eq:drift}) becomes a Dirac function. The singular current sheet 
drift velocity is not physical and is not easy to deal with in the numerical 
method \citep{Zhang1999}. Therefore, we replace the current sheet drift magnitude 
with a formula shown in Equation (\ref{eq:vns}) by following \citet{Burger1989}. 
With the assumption of \citet{Burger1989}, a particle will experience current 
sheet drift if its distance $d$ to the HCS is 
less than two gyro radii $2R_L$, and the magnitude of $\mathbf{v}_{ns}$ is given by
\begin{equation}
v_{ns,0}=vqA\left[0.457-0.412\frac{d}{R_L}+0.0915\left(\frac{d}{R_L}\right)^2
	\right].
\label{eq:vns}
\end{equation}
It can be derived directly from Equation 
(\ref{eq:drift}) that the direction of current sheet drift velocity lies in the HCS and is
perpendicular to the HMF, and the radial, latitudinal, and azimuthal components of the 
current sheet drifts can be written as
\begin{eqnarray}
v_{ns,r} & = & v_{ns,0}\frac{\eta\tan\alpha\cos\phi^\prime\sin\theta^\prime + 
	\Gamma}{\rho}\label{eq:hcsdriftr}\\
v_{ns,\theta} & = & -v_{ns,0}\frac{\tan\alpha\cos\phi^\prime\sin\theta^\prime+
{\Gamma}^2\tan\alpha\cos\phi^\prime\sin\theta^\prime}{\rho}\label{eq:hcsdrifttheta}\\
v_{ns,\phi} & = & v_{ns,0}\frac{1-\eta\Gamma\tan\alpha\cos\phi^\prime\sin\theta^\prime}
	{\rho},
\label{eq:hcsdriftphi}
\end{eqnarray}
respectively, where
\begin{equation}
\phi^\prime=\phi+\frac{\left(r-r_s\right)\Omega}{V_{sw}},
\end{equation}
and
\begin{equation}
\rho=\sqrt{x_1^2+x_2^2+x_3^2}
\end{equation}
with
\begin{eqnarray}
x_1&=&\eta\tan\alpha\cos\phi^\prime\sin\theta^\prime + \Gamma\\
x_2&=&\tan\alpha\cos\phi^\prime\sin\theta^\prime+
{\Gamma}^2\tan\alpha\cos\phi^\prime\sin\theta^\prime\\
x_3&=&1-\eta\Gamma\tan\alpha\cos\phi^\prime\sin\theta^\prime.
\end{eqnarray}
We should note that Equations (\ref{eq:hcsdriftr})-(\ref{eq:hcsdriftphi}) are equal to the 
results of 
\citet{Burger2012} and \citet{Pei2012} if we use the Parker HMF (i.e., $\eta=0$). 
Similar expressions of Equations (\ref{eq:gcdriftr})-(\ref{eq:gcdriftphi}) and 
(\ref{eq:hcsdriftr})-(\ref{eq:hcsdriftphi})
 are given by \citet{Raath2016}, but using different expressions for the HMF.

Since the Parker HMF is an oversimplification and gives a low magnetic field 
intensity at large radial distance, especially at high latitudes,
drifts become very large over the polar regions of the heliosphere. 
It is also known that, with the assumption of weak scattering, the particle's gyro-radius is equivalent  
to its drift scale, so that Figure \ref{fig:RL} also shows that for Parker field
in the polar regions with large solar radial distance, $1$ GV particles' drift speed becomes
very large, but for the modified model the drift speed keeps in the similar level as in 
other regions.

\subsection{Magnetic turbulence throughout the inner-heliosphere}

The development of magnetic turbulence transport models 
\citep[TTMs, see, e.g., ][]{Zank1996,Zank2012,Zank2017,Breech2008,Pei2010a,Oughton2011,
Engelbrecht2013a,Guo2016}, which allows us to have a better scenario of the heliosphere, plays an 
important role in the ab initio models  for cosmic ray diffusion. Especially in the numerical
modulation study, the analytical expressions of diffusion is essential, which
are directly based on the spatial dependence of magnetic turbulence.

Some theoretical work has been done to study TTMs. For example,
\citet{Oughton2011} developed the two-component TTM. Furthermore, 
\citet{Engelbrecht2013a,Engelbrecht2013b} solved the TTM of \citet{Oughton2011} 
for solar minimum interplanetary conditions and it was shown that the results of 
turbulence quantities agree well with the observations of Ulysses, and consequently 
they studied the spatial variations of diffusion coefficients by applying 
the results of TTM in scattering theory, and the results were used in an ab 
initio model for cosmic ray modulation. Recently, \citet{Zank2017} studied 
turbulence quantities with the nearly incompressible magnetohydrodynamics 
(NI MHD) theory. The NI MHD theory can be used to investigate a broader range of 
solar wind observations, and its solutions given by \citet{Adhikari2017} enhance 
our understanding of turbulence quantities in the inner heliosphere. However, 
the TTM from \citet{Engelbrecht2013a} and \citet{Zank2017} are complicated to some extent, 
especially for the study of long-term modulation of GCRs. Therefore, some works 
use analytical expressions to describe the radial and latitudinal dependence of 
magnetic turbulence \citep{Zank1996,Burger2008,Effenberger2012,Ngobeni2014}. 
The magnetic turbulence magnitude in the heliosphere is assumed decreasing as
\citep[e.g., ][]{Burger2008,Effenberger2012,Guo2014,Ngobeni2014,Strauss2017} 
\begin{equation}
\delta B\sim r^S
\end{equation}
where $S$ means the radial dependence of the magnetic turbulence magnitude 
which can be determined later. Based on the observations of Ulysses instruments 
\citep{Perri2010}, the variance of magnetic field magnitude at high latitude is 
smaller than that of low latitude, we can give an expression to describe the 
latitudinal dependence of the magnetic turbulence magnitude as
\begin{equation}
\delta B\sim \frac{1+\sin^2\theta}{2}.
\end{equation}
Therefore, the analytical expression of magnetic turbulence magnitude can be 
written as
\begin{equation}
\delta B = \delta B_{1 \mathrm{AU}}R^S\left(\frac{1+\sin^2\theta}{2}\right),
\label{eq:turbulence}
\end{equation}
where $\delta B_{1 \mathrm{AU}}$ represents the observation of turbulence at 
Earth and $R=r/r_0$, $r_0=1$ AU.  

\begin{figure}
\epsscale{1.} \plotone{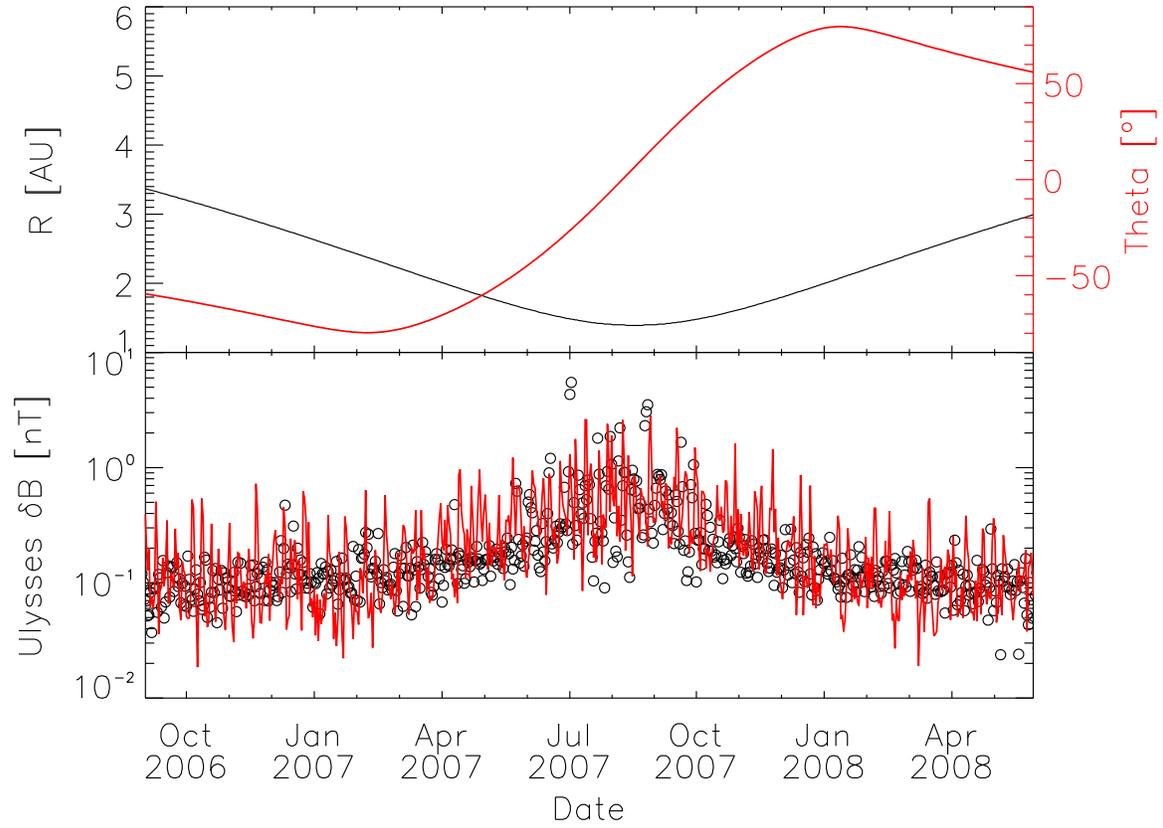}
   \caption{Comparison of the turbulence model with the observation data from 
Ulysses during the Ulysses fast latitude scan in 2007. Top panel shows the 
radial distance and heliographic latitude of Ulysses. Black circles in the 
bottom panel mean the square root of magnetic field variance which are computed 
over 1 day intervals using hourly magnetic field data of Ulysses. 
Red line in the bottom panel represents the result of TRST model. 
$\delta B_{1AU}$ is calculated in the same way using magnetic field data of OMNI.}
   \label{fig:ulyssesdb}
 \end{figure}

According to the observations of Ulysses, $S$ should be varying as a function
of time \citep{Smith2008}, so it is assumed that $S$ is closely associated 
with solar activities. Obviously, there exist different expressions to 
describe such relationship. However, in this work we use the following 
expression,
\begin{equation}
S=-1.56+0.09\ln\frac{\alpha}{\alpha_c},
\label{eq:S}
\end{equation}
where $\alpha_c=1^\circ$. Hereafter we denote the turbulence magnitude 
model from Equations (\ref{eq:turbulence}) and (\ref{eq:S}) as TRST 
(Turbulence magnitude varying as R, S, and Theta) model.
Next, it can be shown that the TRST model agrees well with the turbulence magnitude
measurements from Ulysses, Voyager 1, and Voyager 2, 

Figure \ref{fig:ulyssesdb} represents the comparison between the results of our 
TRST model and the observations of Ulysses 
from September 2006 to May 2008. Top panel shows the trajectory of Ulysses, 
with black and red lines representing the radial distance and heliographic 
latitude, respectively. During this time period, Ulysses had a fast 
latitude scan, the radial distance varies from $1.4$ AU to $3.4$ AU and the 
latitude varies from $-80^\circ$ to $80^\circ$. 
In the bottom panel, black circles
indicate the square root of magnetic field variance which are computed over 1 day 
intervals using hourly magnetic field data of Ulysses, red line represents the 
results of the TRST model, and $\delta B_{1AU}$ is calculated in the same way 
using magnetic field data of OMNI. Considering the 
time delayed heliosphere, the Ulysses data has been shifted back to $1$ AU with 
the typical solar wind speed $v_0=0.25$ AU/day. Note that we 
compute the magnetic field variance over 1 day intervals instead of Carrington 
rotation intervals due to the large latitude variation during the fast latitude 
scan. From this figure we can see that the turbulence model TRST provides
a good prediction of $\delta B$ observations of Ulysses. 

\begin{figure}
\epsscale{1.} \plotone{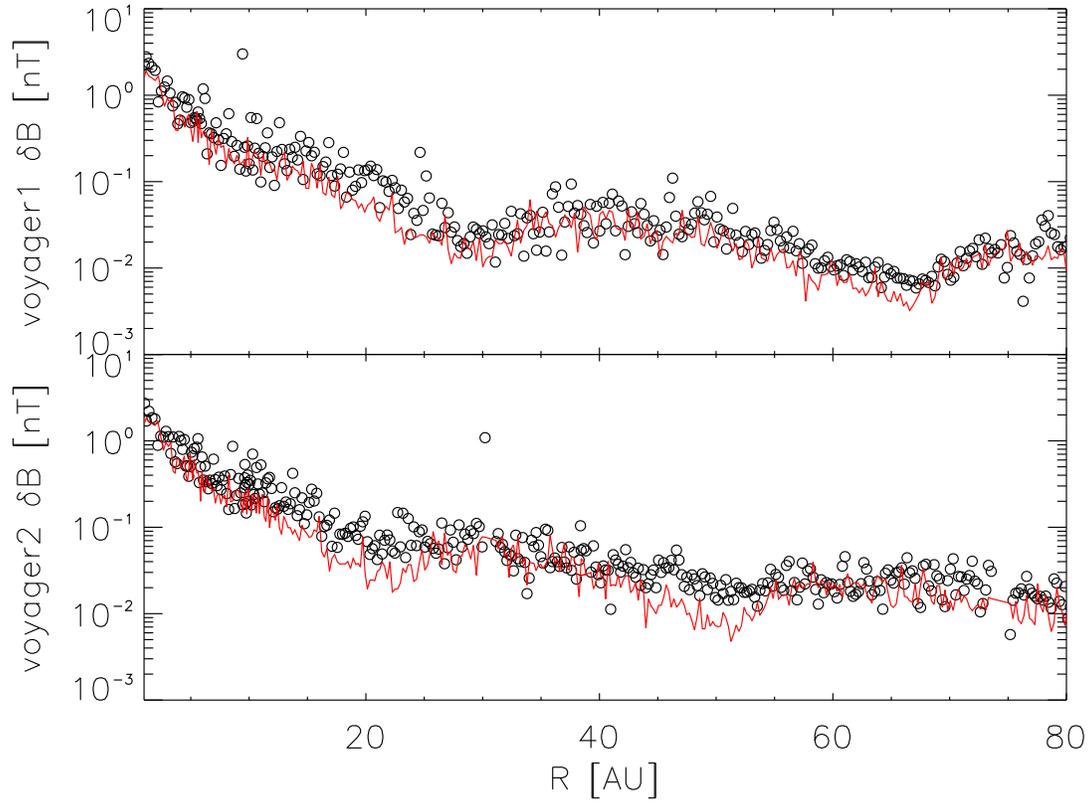}
   \caption{Comparison of the turbulence model TRST results with the observation data 
   from Voyager 1 (top) and Voyager 2 (bottom). 
Black circles indicate the square root of magnetic field variance 
computed over Carrington rotation intervals using hourly magnetic field data 
of Voyager 1 (top) and Voyager 2 (bottom). Red lines represent the results 
of TRST, and $\delta B_{1AU}$ is calculated in the same way using magnetic 
field data of OMNI.}
   \label{fig:voyagerdb}
 \end{figure}

Figure \ref{fig:voyagerdb} shows magnetic turbulence as a function of radial 
distance from $2$ AU to $80$ AU. Black circles in the top and bottom 
panels mean the square root of magnetic field variance which are computed over 
Carrington rotation intervals using hourly HMF data of Voyager 1 and Voyager 2, 
respectively. We have also computed the magnetic field variance 
using methods referred by other works, e.g.,
\citet{Zank1996,Smith2006,Isenberg2010,Adhikari2015}, 
and the results show little difference. Considering the time delayed 
heliosphere, the Voyager data has 
been shifted back to $1$ AU with the typical solar wind speed $v_0=0.25$ AU/day. 
Red lines indicate the results of the TRST model, with $\delta B_{1AU}$ 
calculated in the same way using hourly HMF data from OMNI. It is shown that 
the TRST model provides a good prediction of the Voyager 1 and 2 observations. 
From the top panel we can see that $\delta B$ of Voyager 1 decreases faster during 
solar minimum to form wave troughs, but from  bottom panel the data of Voyager 2
does not show clear wave troughs. It is assumed that the difference between the 
Voyager 1 and 2 magnetic variance is caused by the latitudinal dependence of
turbulence magnitude. Generally speaking, the model TRST provides a 
good prediction of the turbulence variation properties in the inner heliosphere.

Here, we consider two-component model of turbulence \citep{MatthaeusEA90} in solar wind.
It is also necessary for one to know the transport of other properties of turbulence in addition
to magnitude. However, for simplicity, we assume only the turbulence magnitude is varying.

\subsection{Diffusion coefficients}

Turbulent magnetic fields in the solar wind plasma result in diffusion of the 
cosmic rays parallel and perpendicular to the background HMF,
which plays an important role in the modulation processes. 
In the scattering theory we usually emphasis on the global behavior of 
diffusion coefficients (or mean free paths). In the field-aligned 
coordinate, the symmetric and diagonal diffusion tensor $\mathbf{K}_s$ is 
composed of three parts: a parallel diffusion coefficient $\kappa_\parallel$ 
and two perpendicular diffusion coefficients, $\kappa_{\perp r}$  and 
$\kappa_{\perp\theta}$, the perpendicular diffusion coefficients in the 
radial and polar directions, respectively. 
\citet{Jokipii1966} developed the quasi-linear theory (QLT) of the diffusion 
of cosmic rays, which is considered one of the milestones of
the study of cosmic rays. It has been considered that QLT is relatively good to describe
the parallel diffusion of cosmic rays, but perpendicular diffusion has long
been a puzzle. Therefore, empirical models are used in many studies. 
For example, in some study of GCR modulations, empirical expressions for the parallel 
diffusion coefficient are used based on
the QLT, and perpendicular ones are set to be proportional to the parallel one
\citep[e.g.,][]{Potgieter2013,Potgieter2014,Zhao2014, Potgieter2015,Vos2015,Raath2016,
Guo2016}.
The results with this approach are usually consistent with the observations at 1 AU, 
but some parameters in the diffusion coefficient models have to be decided by comparing 
numerical results with the observations.  

\citet{Matthaeus2003} developed a nonlinear guiding center (NLGC) theory for
perpendicular diffusion which agrees well with numerical simulations. Further, 
\citet{Qin2007} extended the NLGC to describe the parallel diffusion, noted as
NLPA. With the combination of the NLGC and NLPA models one gets
two implicit integral equations which can be solved simultaneously to obtain the
perpendicular and parallel diffusion coefficients. \citet{Qin2014} further improved
the combination of the NLGC and NLPA with some slight modification, then they 
obtained a new model, NLGCE-F, by fitting the 
numerical solution from the improved combination of NLGC and NLPA with polynomials. 
The model NLGCE-F allows one to calculate diffusion
coefficients directly without the iteration solution of integration equations set. 
It is noted that in order to use this model the properties of HMF and turbulence in 
solar wind are necessary.
In this study, we assume that $\kappa_{\perp r}=\kappa_{\perp\theta}$. 
The expressions for NLGCE-F are as follows:
\begin{equation}
\ln\frac{\lambda _\sigma}{\lambda_{slab}} =\sum\limits_{i = 0}^{n_{\sigma 1}} 
a_i^\sigma\left(\ln\frac{R_L}{\lambda_{slab}}\right)^i
\label{eq:diffusion}
\end{equation}
with
\begin{eqnarray}
a_i^\sigma  &=& \sum\limits_{j=0}^{n_{\sigma 2}}b_{i,j}^\sigma\left(\ln\frac{E_{slab}}
{E_{total}}\right)^j\\
b_{i,j}^\sigma&=&\sum\limits_{k = 0}^{n_{\sigma 3}}c_{i,j,k}^\sigma \left(\ln
\frac{\delta B^2}{B^2} \right)^k\\
c_{i,j,k}^\sigma&=&\sum\limits_{l=0}^{n_{\sigma 4}}d_{i,j,k,l}^\sigma 
\left(\ln\frac{\lambda_{slab}}{\lambda_{2D}}\right)^l,
\end{eqnarray}
where $\sigma$ indicates $\perp$ or $\parallel$, $\lambda_\sigma=\frac{3}{v}\kappa_\sigma$, 
$R_L$ means the gyro-radius of 
the particle, $\lambda_{slab}$ and $\lambda_{2D}$ are the spectral bend-over 
scales of the slab and 2D components of turbulence, respectively, 
$E_{total}=\left\langle{\delta B^2}\right\rangle$ and 
$E_{slab}=\left\langle{\delta B^2}_{slab}\right\rangle$ are the magnetic 
turbulence energy from all components and from slab component, respectively, and 
$\delta B/B$ is the turbulence level. 
The coefficients $d_{i,j,k,l}^\sigma$ and polynomial order $n_{\sigma i}$ are provided
by \citet{Qin2014}, and the computer code with parameters for NLGCE-F can be downloaded in
\url{www.qingang.org.cn/code/NLGCE-F}.
From \citet{Qin2014} it is also noted that the model NLGCE-F is valid with the parameters 
\begin{eqnarray}
1&\lesssim&\frac{\lambda_{slab}}{\lambda_{2D}}\lesssim 10^3,\\
10^{-3}&\lesssim&\frac{E_{slab}}{E_{total}}\lesssim 0.85,\\
10^{-4}&\lesssim&\frac{b^2}{B^2}\lesssim 10^2,\\
10^{-5}&\lesssim&\frac{R_L}{\lambda_{slab}}\lesssim 6.3.
\end{eqnarray}
In this work we set 
${\lambda_{slab}}/{\lambda_{2D}}=10.0$ \citep{Matthaeus2003}, 
$\lambda_{slab}=0.02r$ with $r$ being the solar
distance, and $E_{slab}/E_{total}=0.2$ \citep{Bieber1994} throughout the 
heliosphere. The turbulence parameters in solar wind, such as
$\lambda_{slab}$ and $\lambda_{2D}$, can only be observed by spacecraft indirectly with
complicated theoretical study \citep[e.g.,][]{MatthaeusEA90, Adhikari2017, Zank2017}, 
so for simplicity purpose we
set them in simple forms according to some study for solar wind in 1 AU
\citep[e.g.,][]{Matthaeus2003}. 
It is noted that if the particle's energy is not much more than $10$ GeV and the radial 
distance is not larger than the distance of termination shock, the values of input 
parameters in Equation (\ref{eq:diffusion}) are in the ranges of validation.

Using diffusion model Equation (\ref{eq:diffusion}) with the turbulence model
Equation (\ref{eq:turbulence}), we are able to establish a 
time-dependent diffusion coefficients model with all input parameters obtained from the 
spacecraft observations near Earth. 
\citet{Manuel2014} also established a time-dependent diffusion model 
with the time-dependent parameters scaled by HMF magnitude (B) and variance 
($\delta B^2$) \citep[see also,][]{Ferreira2004, Manuel2011,  Potgieter2014}.

\begin{figure}
\epsscale{0.8} \plotone{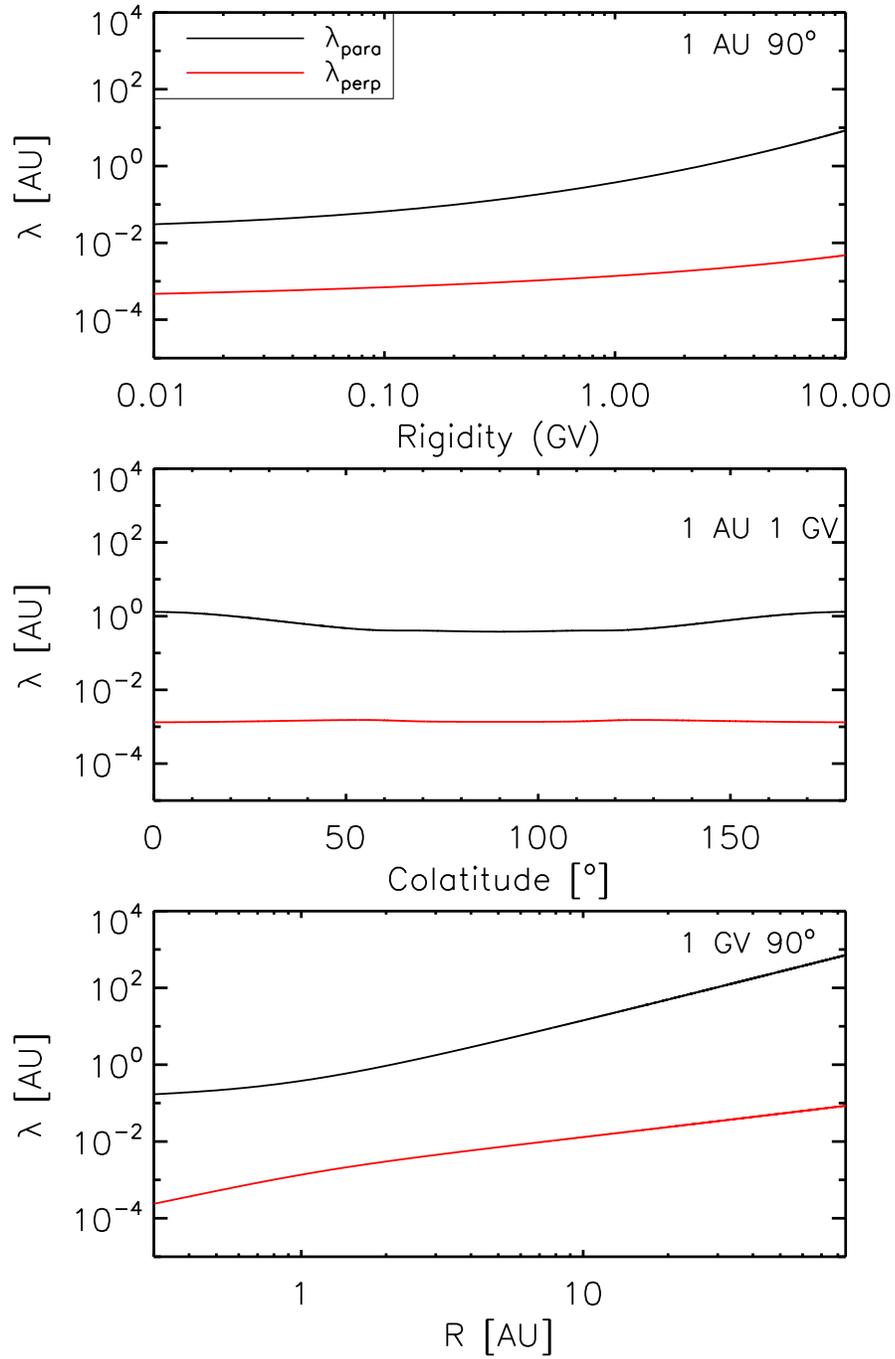}
   \caption{Parallel and perpendicular mean free paths are shown as functions of 
   rigidity, colatitude at $1$ AU, and radial distance in the ecliptic plane in top,
   middle, and bottom panels, respectively. }
   \label{fig:lambda}
 \end{figure}

Figure \ref{fig:lambda} shows scenarios of mean free paths as a function of 
rigidity (top panel), polar angle (second panel), and radial distance 
(bottom panel). Interplanetary conditions are the same as
that in Figure \ref{fig:RL}, and $\delta B_{1AU}=3.0$ nT. 
The parallel mean free path $\lambda_\parallel$ shows stronger rigidity 
dependence when the rigidity increases, and $\lambda_\parallel$ is larger in 
the polar region as have been shown in the first and second panels. 
At lower energy, i.e., when energy is lower than about 3 GV, the parallel 
mean free path shows the expected $P^{1/3}$ dependence, but at higher energy the 
model gives a $\sim P^{3/2}$ dependence. The perpendicular mean free path is 
relatively flat as a function of rigidity and colatitude. As has been shown in 
the bottom panel, the parallel and perpendicular mean free paths show a gradual 
increase with the radial distance. 

\subsection{Heliospheric boundary}
Voyager 1 crossed the heliopause at $121.7$ AU in August 2012 and has reached 
the very local interstellar medium, \citet{Zhang2015} believed that the solar 
modulation boundary is located a fraction of an AU beyond the heliopause. 
Therefore, the very local interstellar \citep[LIS, e.g.,][]{Potgieter2014, Vos2015} 
spectrum of GCRs could be used as the input spectrum in our modulation model. 
However, in our model, we set the outer boundary at a smaller solar distance $r=85$ AU,
for simplicity purpose, so that we do not include the termination shock
acceleration of GCRs and other complicated phenomenon in outer heliosphere. 
Furthermore, to keep solar distance $r$ not too large could make sure the 
diffusion model NLGCE-F always valid in this work. In addition, we assume the GCR
source at $r=85$ AU as
\begin{equation}
j_{S}=J_0p_0^{2.6} p\left({m_0}^2c^2+p^2\right)^{-1.8}
\label{eq:lis}
\end{equation}
where $J_0$ is a constant determined later and 
$p_0=1~\mathrm{GeV}/c$ by following \citet{Zhang1999}.

\section{Numerical Methods}
To solve the Parker transport equation, we make use of the time-backward Markov 
stochastic process method proposed by \citet{Zhang1999}. For a pseudo-particle 
in position $(r,\theta,\phi)$ and momentum $p$, the stochastic differential 
equations equivalent to Equation (\ref{eq:TPE}) have the form 
\citep{Zhang1999,Pei2010,Strauss2011,Kopp2012}
\begin{equation}
dx_i=A_i(x_i)ds+\sum_{j}B_{ij}(x_i) \cdot dW_j,
\end{equation} 
with $i\in{(r,\theta,\phi,p)}$, $x_i$ the Ito processes \citep{Zhang1999}, 
$s$ the backward time and $\mathrm{d}W_i$ satisfy a Wiener process given by the 
standard normal distribution \citep{Pei2010,Strauss2011}. 
For the modified Parker HMF used in this work, the matrix components $B_{ij}$ 
are given by \citet{Pei2010} \citep[see also][]{Kopp2012}, 
\begin{eqnarray}
B_{11}&=&\sqrt{\frac{2(\kappa_{\phi\phi}\kappa_{r\theta}^2-2\kappa_{r\phi}\kappa_{r\theta}\kappa_{\theta\phi}+\kappa_{rr}\kappa_{\theta\phi}^2+\kappa_{\theta\theta}\kappa_{r\phi}^2-\kappa_{rr}\kappa_{\theta\theta}\kappa_{\phi\phi})}{\kappa_{\theta\phi}^2-\kappa_{\theta\theta}\kappa_{\phi\phi}}}\\ 
B_{12}&=&\frac{\kappa_{r\phi}\kappa_{\theta\phi}-\kappa_{r\theta}\kappa_{\phi\phi}}{\kappa_{\theta\phi}^2-\kappa_{\theta\theta}\kappa_{\phi\phi}}\sqrt{2\left(\kappa_{\theta\theta}-\kappa_{\theta\phi}^2/\kappa_{\phi\phi}\right)}\\
B_{13}&=&\kappa_{r\phi}\sqrt{\frac{2}{\kappa_{\phi\phi}}} \\
B_{22}&=&\frac{1}{r}\sqrt{2\left(\kappa_{\theta\theta}-\kappa_{\theta\phi}^2/\kappa_{\phi\phi}\right)}\\
B_{23}&=&\frac{\kappa_{\theta\phi}}{r}\sqrt{\frac{2}{\kappa_{\phi\phi}}}\\
B_{33}&=&\frac{\sqrt{2\kappa_{\phi\phi}}}{r\sin\theta}\\
B_{21}&=& B_{31}=B_{32}=0,
\label{eq:matrixB}
\end{eqnarray} 
and the components of vector $\mathbf{A}$ are given as follows, 
\begin{eqnarray}
A_r&=&\frac{\partial \kappa_{rr}}{\partial r}+\frac{2}{r}\kappa_{rr}+\frac{1}{r}\frac{\partial \kappa_{r\theta}}{\partial\theta}+\frac{\cot\theta}{r}\kappa_{r\theta}+\frac{1}{r\sin\theta}\frac{\partial \kappa_{r\phi}}{\partial\phi}-V_{sw}-v_{d,r}\\
A_\theta&=&\frac{1}{r}\frac{\partial \kappa_{r\theta}}{\partial r}+\frac{1}{r^2}\kappa_{r\theta}+\frac{1}{r^2}\frac{\partial \kappa_{\theta\theta}}{\partial\theta}+\frac{\cot\theta}{r^2}\kappa_{\theta\theta}+\frac{1}{r^2\sin\theta}\frac{\partial \kappa_{\theta\phi}}{\partial\phi}-\frac{1}{r}v_{d,\theta}\\
A_\phi&=&\frac{1}{r\sin\theta}\frac{\partial \kappa_{r\phi}}{\partial r}+\frac{1}{r^2\sin\theta }\kappa_{r\phi}+\frac{1}{r^2\sin\theta}\frac{\partial \kappa_{\theta\phi}}{\partial\theta}+\frac{1}{r^2\sin^2\theta}\frac{\partial \kappa_{\phi\phi}}{\partial\phi}-\frac{1}{r\sin\theta}v_{d,\phi}\\
A_p&=&\frac{p}{3r^2}\frac{\partial r^2V_{sw}}{\partial r}.
\label{eq:vectorA}
\end{eqnarray} 
Therefore the statistical differential equations can be written as 
\begin{eqnarray}
\mathrm{d}r & = & A_r\mathrm{d}s+B_{11}\mathrm{d}W_r+B_{12}\mathrm{d}W_\theta+B_{13}\mathrm{d}W_\phi \label{eq:SDEr}\\
\mathrm{d}\theta &=& A_\theta\mathrm{d}s+B_{22}\mathrm{d}W_\theta+B_{23}\mathrm{d}W_\phi \label{eq:SDEtheta}\\
\mathrm{d}\phi & =& A_\phi\mathrm{d}s+B_{33}\mathrm{d}W_\phi \label{eq:SDEphi}\\
\mathrm{d}p & = & A_p\mathrm{d}s. \label{eq:SDEp}
\end{eqnarray}
 
Note that the diffusion tensor in Equations (\ref{eq:SDEr}-\ref{eq:SDEp}) 
are elements of the symmetric diffusion tensor $\mathbf{K}_s$ in spherical coordinates. 
According to \citet{Burger2008} elements of $\mathbf{K}_s$ in spherical 
coordinates for the modified Parker HMF are written as 
\begin{eqnarray}
\kappa_{rr}&=&\kappa_{\perp\theta}\sin^2\zeta + \cos^2\zeta(\kappa_{\parallel}\cos^2\Psi 
+ \kappa_{\perp r}\sin^2\Psi) \\ 
\kappa_{r\theta}&=&\kappa_{\theta r}=\sin\zeta\cos\zeta(\kappa_{\parallel}\cos^2\Psi + 
\kappa_{\perp r}\sin^2\Psi - \kappa_{\perp\theta})\\
\kappa_{r\phi}&=&\kappa_{\phi r}= - (\kappa_{\parallel} - 
\kappa_{\perp r})\sin\Psi\cos\Psi\cos\zeta \\
\kappa_{\theta\theta}&=&\kappa_{\perp\theta}\cos^2\zeta + 
\sin^2\zeta(\kappa_{\parallel}\cos^2\Psi + \kappa_{\perp r}\sin^2\Psi) \\
\kappa_{\theta\phi}&=&\kappa_{\phi\theta}=- (\kappa_{\parallel} - 
\kappa_{\perp r})\sin\Psi\cos\Psi\sin\zeta \\
\kappa_{\phi\phi}&=&\kappa_{\parallel}\sin^2\Psi + \kappa_{\perp r}\cos^2\Psi, 
\end{eqnarray}
with $\tan\Psi=-B_{\phi}/(B_r^2 + B_{\theta}^2)^{\frac{1}{2}}$ and 
$\tan\zeta=B_{\theta}/{B_r}$, where $\Psi$ is the HMF winding angle. 

In our modulation model we use the time-delayed interplanetary 
conditions at radius $r$ related to the states at the source surface $r_s$ 
at some earlier time, and the heliosphere is considered dynamic due to 
the solar activities.  

\section{Modeling Results}

\begin{figure}
\epsscale{1.} \plotone{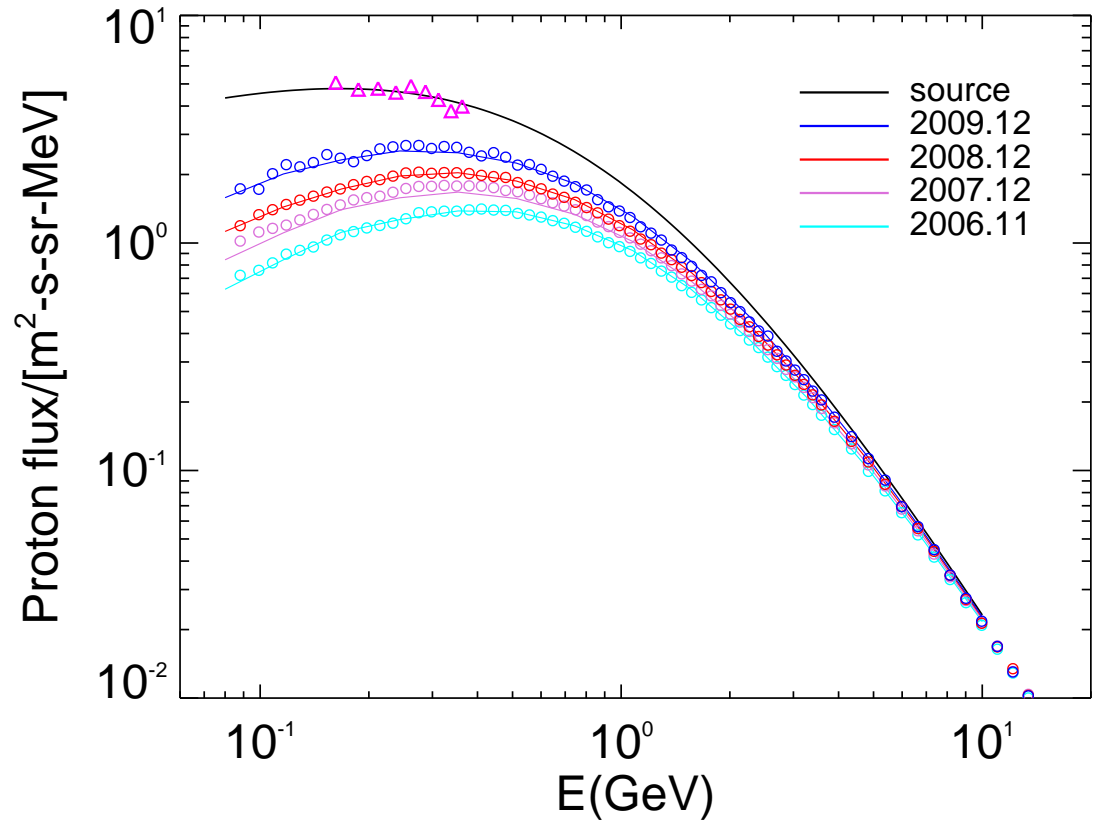}
   \caption{Computed GCR energy spectra at Earth for the period from 2006 to 
2009 (color lines). Circles are observations of the PAMELA instrument. 
Black line means the GCR source at $85$ AU, and magenta triangles represent 
Voyager 2 observations at 85 AU reported by \citet{Webber2008}.}
   \label{fig:pamela1}
 \end{figure}

In our GCR modulation model we only need four input parameters which can be 
obtained from the observations at $1$ AU, i.e., the heliospheric current 
sheet tilt angle, the solar wind speed, the magnitude of background magnetic 
field $B$, and the magnetic turbulence magnitude $\delta B_{1AU}$. 
Figure \ref{fig:pamela1} shows the computed and observed proton spectra for 
four Carrington rotations in November 2006, December 2007, December 2008, and 
December 2009 with colors cyan, purple, red, and blue, respectively. 
Circles means observations of PAMELA, numerical results of modulation modeling are 
shown as solid lines. The GCR source at $85$ AU 
is represented by the black line with the constant $J_0=1.17\times 10^4$
$\mathrm{m}^{-2}\mathrm{s}^{-1}\mathrm{sr}^{-1}\mathrm{(GeV/nuc)}^{-1}$ in Equation
(\ref{eq:lis}), and magenta triangles mean Voyager 2 
observations at 85 AU reported by \citet{Webber2008}. The modulation results 
show good agreement with the observations of PAMELA.

\section{Discussion and Conclusions} 

In this work, we develop a numerical model to study the time-dependent 
modulation of cosmic rays in recent solar minimum with PAMELA observations. 
We use the time-backward Markov stochastic process method \citep{Zhang1999} to 
numerically solve the Parker transport equation. In our GCR modulation model, 
all the parameters are obtained from the observations of OMNI.
We get galactic proton spectra varying as a function of time during the recent 
solar minimum, which are consistent with the observations of PAMELA. 

As the Parker HMF provides a low magnitude in the polar regions at large 
radial distance, we adopt the modified Parker HMF according to 
\citet{Jokipii1989}, which can help to avoid the traditional Parker HMF's
problem that the drift speed of GCRs in polar regions are too large. 
Considering the dynamic phenomena of heliosphere with the solar wind flow
from source surface to any solar distance, we use the input parameters observed
near the Earth in the earlier time with a typical solar wind speed $v_0=0.25$ 
AU/day. It is noted that this will divide the heliosphere into slices and 
introduce an additional radial dependence in the HMF magnitude, i.e., 
$B_0=B_0(r)$. In the region
between two slices of plasma the HMF is not divergence free, but in each step of the
numerical solution of the TPE Equation (\ref{eq:TPE}), we always keep inside one slice of
plasma. In addition, 
we set the outer boundary at a smaller solar distance $r = 85$ AU, for simplicity 
purpose, so that we do not include the termination shock acceleration of GCRs 
and other complicated phenomenon in outer heliosphere. The GCR source spectrum we use 
is consistent with the observations of Voyager 2 at 85 AU reported by \citet{Webber2008}. 
Furthermore, by keeping solar 
distance $r$ not too large we can make sure the diffusion model
NLGCE-F always valid in this work. 

The knowledge of transport of magnetic turbulence
throughout the heliosphere is very important to determine the diffusion 
coefficients. According to previous studies 
\citep[e.g., ][]{Zank1996,Burger2008,Effenberger2012,Ngobeni2014,Strauss2017,
Perri2010}, we use a model for magnetic turbulence magnitude with 
Equation (\ref{eq:turbulence}), i.e., $\delta B\sim r^S(1+\sin^2\theta)$,
in which the latitudinal dependence is assumed from the observations of Ulysses, 
and the expression of $S$ for the radial dependence is chosen as a function of 
the heliospheric current sheet tilt angle with Equation (\ref{eq:S}). We show 
that the new turbulence magnitude model with Equations (\ref{eq:turbulence}) 
and (\ref{eq:S}), denoted as TRST model, agrees well with the Ulysses, Voyager 1, 
and Voyager 2 observations. In addition, we assume two-component model 
of turbulence in solar wind. For simplicity purpose, we only suppose the 
magnetic turbulence magnitude is varying. 

We use the new diffusion model NLGCE-F from \citet{Qin2014} which
was obtained by fitting the numerical solution from the non-linear parallel and 
perpendicular diffusion with polynomials. The using of the diffusion model NLGCE-F
helps us to get more accurate diffusion coefficients without consuming lots of 
computing resources.
For the drift coefficient, turbulence can provide the suppression 
 \citep[see, e.g., ][]{Jokipii1993,Fisk1995,Giacalone1999,Candia2004,Stawicki2005,
Minnie2007,Tautz2012}. 
The reduction of drift effects is complicated to be used self-consistently 
\citep[see, e.g.,][]{Bieber1997, Burger2010, Tautz2012} or in Ad hoc form 
\citep[see, e.g.,][]{Burger2000,Potgieter2013,Vos2016,Nndanganeni2016} in 
modulation works. It is far from complete to understand the effects 
of turbulence on CR drifts. Therefore, in this work, we use the weak scattering 
drift coefficient for simplicity purpose. 

In the future, we plan to use the modulation model established in this paper to study 
the 11 and 22 year modulation of GCRs in the 
inner heliosphere \citep[e.g.,][]{McDonald1998, ShenAQin2016}. If our model works
well, we can 
reproduce the GCR observations by Ulysses, Voyager 1, and Voyager 2 with long
period of time. Otherwise, we need to improve our modulation model. Firstly, we 
could improve the turbulence model by modifying the magnitude model and applying
more realistic models for transport of turbulence geometry. 
Secondly, we could use a more self-consistent
dynamic heliosphere model, e.g., a model from MHD simulation. 
Thirdly, we could include termination shock in 
the model to study the realistic boundary effects. Fourthly, the GCR source spectrum
could be improved. Fifthly, the drift suppression from turbulence could be
included.

\acknowledgments
We are partly supported by grants NNSFC 41374177, NNSFC 41574172, 
and NNSFC 41125016, and the Specialized Research Fund for State Key 
Laboratories of China. We used the HCS tilt angle data from the Wilcox Solar 
Observatory ( \url{wso.stanford.edu} ), solar wind speed and magnetic field data 
from OMNI website ( \url{https://omniweb.gsfc.nasa.gov/} ), and PAMELLA data from 
Database  for  Charged  Cosmic  Ray  measurements 
( \url{https://tools.asdc.asi.it/CosmicRays/} ). 
The work was carried out at National Supercomputer Center in Tianjin, 
and the calculations were performed on TianHe-1 (A).

\clearpage

\clearpage

\clearpage

\clearpage

\clearpage

\clearpage

\clearpage

\clearpage
\end{document}